\documentstyle[prb,epsf,psfig,twocolumn,aps]{revtex}
\begin{document}
\twocolumn[\hsize\textwidth\columnwidth\hsize\csname@twocolumnfalse%
\endcsname
\title{Straight cracks in dynamic brittle fracture}
\author{O. Pla$^1$, F. Guinea$^1$, E. Louis$^2$, S.V. Ghaisas$^3$, and 
L. M. Sander$^4$} \address{
$^1$ Instituto de Ciencia de Materiales, Consejo Superior de Investigaciones
Cient{\'\i}ficas, Cantoblanco, E-28049 Madrid, Spain. \\
$^2$ Departamento de F{\'\i}sica Aplicada, Universidad de Alicante,
Apartado 99, E-03080 Alicante, Spain. \\
$^3$ Department of Electronics Science, University of Pune, Pune 411007, India 
\\
$^4$
Physics Department, The University of Michigan, Ann Arbor MI 48105-1120. USA}

\date{\today}

\maketitle

\begin{abstract}
We study the dynamics of cracks in brittle materials when the velocity
of the crack is comparable to the sound velocity by means of lattice
simulations. Inertial and damped dynamics are analyzed. It is shown that
dissipation strongly influences the shape of the crack. While inertial
cracks are highly unstable, dissipation can stabilize straight cracks.
Our results can help to explain recent experiments on PMMA.
 
\end{abstract}

]
\narrowtext
\section{Introduction.}
The dynamics of cracks in brittle materials such as glasses has recently
 attracted a great
deal of interest. While an extensive body of work  exists on the 
properties of quasistatic cracks, crack propagation when the crack
grows at velocities comparable to the sound velocity is still
poorly understood (see   \cite{Freund}). Particular attention has been devoted
to the study of crack tip instabilities such as crack branching and
oscillation~\cite{Swinney,Ciliberto1,Ciliberto2,Fineberg1,Fineberg2}. 
Typically, the crack tip reaches a critical velocity of the order of the
Rayleigh speed in the material; faster cracks branch or oscillate. Interesting
patterns were also observed under an applied thermal gradient~\cite{thermal}.

In the following, we analyze crack tip instabilities in brittle
materials. In these systems, the stress distribution around the
crack is assumed to be well described by the continuum theory of
elasticity \cite{Landau}. We assume that the instabilities observed
in the experiments cited above are determined by these stresses.

The stress distribution near the tip of a moving crack was analytically
calculated by Yoffe \cite{Yoffe}. The calculation shows a bifurcation at a
critical velocity $c_Y \approx 0.6 c_R$ where $c_R$ is the Rayleigh velocity.
Beyond this velocity, the stress component tangent to the crack tip
($\sigma_{\theta \theta}(r,\theta )$,
assuming that the tip has radius of curvature, $r$)
has a maximum at a finite angle $\theta$ with the crack direction.
This result can be interpreted as a tendency for the crack tip to
deviate from the straight direction.

This criterion is the simplest which predicts an instability of
an inertial crack. Alternative criteria can be obtained by 
slightly perturbing the crack shape, and looking for the growth of
the perturbation. In\cite{AB96}, a wavy perturbation is added to
a straight, quasistatic crack,
and the induced modifications of the stresses at the tip are calculated.
An instability is identified when the shear stresses are enough
to deviate the crack from its initial directions.
In this quasistatic case, a finite external shear is required
to induce the instability. This analysis is extended
to dynamical cracks in\cite{AABL98}, where it is shown that
above a certain velocity an infinitesimal shear distortion is amplified.
The critical velocity depends on material parameters
which describe the forces at the crack tip. Quasistatic cracks in PMMA
under different stress distributions follow trajectories well
described in terms of the stress intensity factor near the crack
tip\cite{GEGP96}.

We now concentrate on cracks in PMMA,
which is a glassy polymer. The microscopic aspects of the fracture process
are not understood. It is possible to define a characteristic length
in terms of the tensile strength, $f_t$, the crack surface energy,
$G_F$, and the Young's modulus of the material. The standard
definition for thick plates and planar deformations is\cite{thesis}
$l_{\rm ch} = E G_F/((1 - \nu^2 )f_t^2)$. 
This length, derived from macroscopic
parameters, gives an estimate of the scale at which continuum elasticity
may cease to be valid. Using the parameters in Table I, we find:
$l_{ch} \approx
60.2$ microns\cite{thesis}.
Typical structures at the crack tip, such as its radius of curvature,
have dimensions comparable to this length. Thus, a comprehensive model of
the growing crack should, at least, take into account physical phenomena 
beyond this scale. Elastic waves of wavelengths of micrometers have 
frequencies in the gigahertz range. 

\begin{table}
\begin{tabular}{lc}
 $c_e$ (cal g$^{-1}$ K$^{-1})$ & 0.28   \\
$\kappa$ (cal cm$^{-1}$ s$^{-1}$ K$^{-1}$)&4.7$\times 10^{-3}$ \\
 $\rho$ (g cm$^{-3}$) & 1.2  \\
 $1/(\kappa\rho c_e)$  & 6.3$\times 10^{3}$  \\
 $E$ (GPa)  &  2.9   \\
 $\nu$ (Poisson ratio) &  0.401 \\
 $G_F$ (N m$^{-1}$) & 290    \\
 $f_t$ (MPa) & 130    \\
 $c_R$ ( m s$^{-1}$) & 989    \\
\end{tabular}
\caption{
Experimental values for some constants of PMMA relevant
to the present work. The meaning of the symbols is:
$c_e$, specific heat, $\kappa$, thermal conductivity, $\rho$, density, $E$,
Young's modulus, $G_F$, crack surface energy, $f_t$, tensile strength,
and $c_R$, Rayleigh velocity.}
\label{constants}
\end{table}

In the following we will analyze cracks in PMMA by means of an
approximate model in which details at the atomic and molecular
scale are neglected. We take it as a coarse-grained approximation
to a more microscopic description in order to gain
information on the role of effective macroscopic parameters
on the shape of the growing crack. The model has already
been used in studying the influence of thermal gradients
on crack growth\cite{P96}. As we will discuss, we
find that the viscosity (which determines the sound attenuation, for instance)
plays a major role in stabilizing straight cracks and controlling
their instabilities. 

In the following section, we give a brief discussion of the 
experimental situation. Then, we discuss the model, its general features,
and related results available in the literature. We next present 
our results. The following section discusses possible improvements
of the model, and the article ends with a conclusion section.
There is an appendix where possible mechanisms which may lead to
dissipation in PMMA are explored.

\section{Experimental Facts}

The propagation of cracks, under mode I conditions, in PMMA shows different
regimes~\cite{Swinney,Ciliberto1,Ciliberto2,Fineberg1,Fineberg2}. Most
experiments are done in PMMA sheets under uniaxial stress. An initial
straight crack grows with a velocity dependent on the applied stress. Above a
certain threshold, the velocity shows oscillations in time, although the crack
remains straight. At higher values of the average velocities the crack surface 
becomes rough due to the formation of microscopic side branches
(microbranching transition). At even higher velocities, the crack branches
into several paths. The transition from velocity oscillations (and acoustic
emission) to microbranching is accompanied by a discontinuity in the average
velocity. These transitions take place at velocities which are a fraction of
the Rayleigh velocity, $c_R = 989$ m/s. Only a small fraction of the
energy dissipated during the growth process is radiated into sound
waves~\cite{Fineberg2}. Significant heating effects have been
reported~\cite{heating}. When the growing crack is perturbed by means of
external sound sources, many features of the previous picture are
modified~\cite{Boudet1}. The velocity gap at the microbranching transition is
washed out. 

The previous picture also seems to hold for cracks in ordinary
glass~\cite{glass}. Cracks moving at constant speed seem, however,
much more difficult to stabilize in ordinary glass than in PMMA~\cite{Boudet2}.

\section{Dissipation in PMMA}
Elastic waves are attenuated in real materials,
and energy is transferred to degrees of freedom other than those
which describe sound waves. This attenuation can be modeled by adding
a  viscosity term ~\cite{Landau} to the elastic equations of motion
of the form $\eta \nabla^2 \partial_t {\bf u}$ where ${\bf u}$ is
the displacement, and $\eta$ is a viscosity coefficient.
In this long-wavelength limit transverse sound waves
acquire an attenuation $\alpha =\eta k^2/2\rho c_T$ where $\rho$ is
the density and $c_T$ is the transverse sound velocity. Thus
there is a wave-vector at which the attenuation of a wave becomes
comparable to its wavelength, $\alpha \Lambda = \pi\eta k/\rho c_T \sim1$. 
Beyond this scale, sound waves are overdamped, and
the analysis reported in  \cite{Yoffe} certainly needs to be modified.

The influence of a different form of
viscosity on the velocity of straight cracks
was considered in\cite{Langer}. It was found that the presence of
damping at the edges of a type III crack leads to a steady state
velocity which, at high viscosities, is inversely proportional to
the damping coefficient.

The term $\eta \nabla^2 \partial_t {\bf u}$ is thought to
be appropriate at very low frequencies. However, 
in glassy systems the attenuation is a complicated function of
frequency due to the different relaxation processes which
contribute~\cite{FA86}. For example, for PMMA  at high frequencies (several
GHz), $\alpha \Lambda \sim 0.1$~\cite{Jackson,Maris}, and, thus,
$\alpha \propto \omega$. At lower frequencies
(2 MHz), the dependence of $\alpha$ on frequency can be fitted by a power law, 
$\alpha \sim \omega^c$, with $c \sim 1.5 - 1.7$\cite{Mark}. It has been
argued that some relaxation processes freeze below 165K\cite{Per}.
It is likely that, at frequencies $<$ 100 GHz
a dependence other than $\omega^{2}$ may arise.
At sufficiently low temperatures, the situation simplifies somewhat, as the
main modes which contribute to dissipation are better 
understood\cite{WE78}. A microscopic analysis of dissipation processes
at low temperatures, using as input detailed experimental data on
the low energy modes in glassy polymers\cite{Cetal98}, is given in the
Appendix.

\section{General features of cracks in brittle solids}

\subsection{The Equations of Motion}

The equations of motion including viscous terms for an isotropic
medium can be written as:
\begin{eqnarray}
\rho \partial_{tt}{\bf u}&=&\mu \nabla^2{\bf u} + (\lambda + \mu)
{\bf \nabla} \left ({\bf \nabla}{\bf u}\right )+ \nonumber \\
&&\eta \nabla^2\partial_t{\bf u} + (\psi + \eta){\bf \nabla}\left(
{\bf \nabla}\partial_t{\bf u}\right),
\label{e:continuum}
\end{eqnarray} 
where ${\bf u}$ is the displacement field, $\lambda$ and $\mu$
the Lam\'e coefficients and $\psi$ and $\eta$ the corresponding
coefficients in the viscous case. 

\subsection{Attenuation of Elastic Waves}
The relation between the attenuation coefficient and the frequency can be 
derived approximating the solutions of the equations
of motion by a longitudinal plane wave, 
\begin{equation}
{\bf u}({\bf r},t) = {\bf u}_0 {\rm e}^
{{\rm i}[(k+{\rm i}\alpha)x-\omega t]}
\end{equation}
\noindent where ${\bf u}_0=u_0{\hat {\bf \i}}$ is the amplitude of the
wave and ${\hat {\bf \i}}$ the unit vector in the $x$--direction. The
result for the attenuation coefficient is,
\begin{eqnarray}  
\alpha =\omega\sqrt{\frac{\rho}{2}}
\left[\frac{1}{\sqrt{(\lambda+2\mu)^2+
(\psi+2\eta)^2\omega^2}}-\right. \nonumber\\
\;\;\;\left.\frac{\lambda+2\mu}{(\lambda+2\mu)^2+
(\psi+2\eta)^2\omega^2}\right]^\frac{1}{2}
\label{e:atenvso}
\end{eqnarray}
\begin{figure}
\epsfxsize=\hsize 
\centerline{\epsfbox{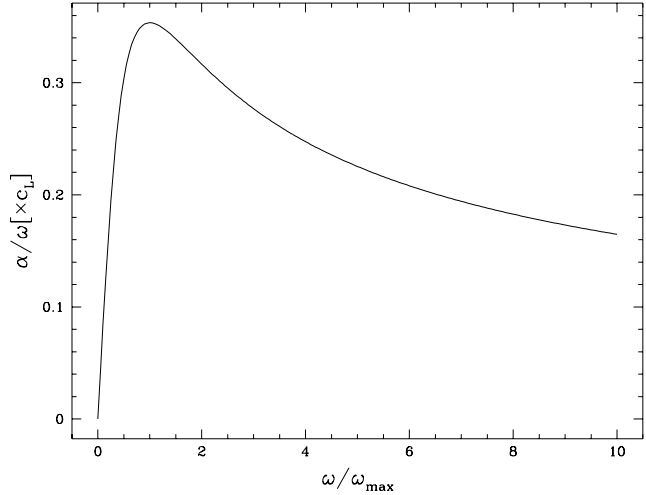}} \vspace{-1.5cm}
\caption{Attenuation coefficient divided by the frequency as function of the
frequency obtained for longitudinal plane waves (see text). The frequency is
expressed in units of the frequency at which the maximum occurs
($\omega_{\rm max}$).}
\label{f:atenvso} 
\end{figure}
The dispersion relation ($k(\omega)$) is just Eq.~\ref{e:atenvso} changing in
the right hand side the minus sign by a plus sign. In Fig.~\ref{f:atenvso} we
plot $\alpha/\omega$ versus $\omega/\omega_{\rm max}$. This ratio has a
maximum at,
\begin{equation}
\omega_{\rm max}=\sqrt{3}\frac{\lambda+2\mu}{\psi+2\eta}
\end{equation}
and the behavior of the attenuation coefficient in the low and
high frequency limits is,
\begin{mathletters}
\begin{equation}
\omega \rightarrow 0, \;\; \alpha \propto \omega^2 
\end{equation}
\begin{equation}
\omega \rightarrow \infty, \;\; \alpha \propto \omega^{1/2}
\end{equation}
\end{mathletters}
\noindent 
These  results show that,
roughly, the attenuation coefficient saturates at wavelengths at which
the attenuation and the frequency of the wave become comparable.
On the other hand, the behavior at low frequencies is consistent with the 
experimental data and with 
the results obtained by means of the microscopic analysis described 
in the Appendix. At high frequencies, however, both the experiments and
the microscopic analysis give $\alpha \propto \omega$. This discrepancy
is not surprising, considering that the continuum theory should fail
at small length scales (see above). 
 
\subsection{Generalization of the Griffith criterion}
The Griffith criterion is a fundamental element in the
theory of fracture \cite{Gr20}. According to Griffith, a
crack starts to advance if, in increasing its length by $\delta L$, 
the elastic energy released is greater than the amount of energy needed to 
create the new fracture surface.
Mott was the first to include kinetic effects in Griffith's analysis 
\cite{Gr20,Mo48}.
He proposed to add a kinetic energy to the Griffith's total energy. 
However some of the conclusions inferred from Mott's analysis are not valid,
for instance the predicted value for the maximum crack speed was lower
than expected (see \cite{Freund} for
a full discussion). 

Here use a different approach and attempt to directly
generalize the Griffith criterion to the viscous case by keeping
track of the energy flows. We balance the energy release, the 
difference of the elastic energy and the surface energy,
with losses due to viscous dissipation. We consider a long system of 
width $W$ and thickness $d$, with a crack of length $L$. 
To estimate the elastic energy release $E_r$, we note that for $L << W$
a roughly round region
of diameter $L$ is fully relaxed, so that $E_r \propto \epsilon^2 L^2 d$
where $\epsilon$ is the strain that causes the material to break.
For $L>>W$ we must put $E_r \sim \epsilon^2 WLd$. The second term,
the cost of creating new fracture surface $E_f$, is $E_f \propto  Ld$. 
Finally, the dependence of the rate of viscous dissipation $\delta E_d$ 
on the crack speed $v$, may be estimated for slow cracks from a 
symmetry argument:
Since $\delta E_d \rightarrow 0$ as $v \rightarrow 0$, and
must be non-negative for any $v$, we conclude that  $\delta E_d
\propto v^2 \delta t$. The coefficient of this term goes to zero as 
$\eta \rightarrow 0$, so that we put $E_d \propto \eta v^2 \delta t$. 
Now we set $\delta E_d = \delta E_r - \delta E_f$, and use 
$\delta L = v \delta t$. For short cracks we see that 
\begin{equation}
\eta v \propto \epsilon^2 L - q \label{eqacc}
\end{equation} 
where $q$ is a constant. Short cracks accelerate. For long cracks, on the
other hand, there is a  \emph{terminal velocity}:
\begin{equation} 
\eta v \propto	\epsilon^2 W - q.\label{eqterm}
\end{equation}
If the terminal velocity is less than
$c_Y$, the Yoffe threshold, we may expect that the 
crack will never branch. 

Note that the analysis of this section can be expected to
be valid only in the limit of low crack speed. In particular,
the viscous dissipation term on the right hand side of Eq.~\ref{eqterm}
involves the motion of the lattice in response to a passing crack.
Thus the $\eta$ in this equation is not quite the same
as the one above, and for substantial $v$ probably has a complicated dependence
on the microscopic $\eta$ and on $v$ itself. We will test these ideas 
with simulations,
below, and find that for speeds $<< c_R$ Eq.~\ref{eqterm} is rather well
obeyed, but that there are deviations at large speed.

\subsection{The branching instability}

The assumption in the previous section is that the 
critical speed for branching is
independent of $\eta$, which is what we find numerically for
small $\eta$ (see below). 
This is a bit unexpected since, in the presence of dissipation 
the analytical solution of Yoffe, for example,
\cite{Yoffe} is no longer correct. We can see where this assumption
would break down by examining the form of that solution. 

The stress field
described can be derived from an appropriate distribution of
forces applied at the crack edges which have the general form ${\bf
 {f}} ( {\bf  {r}} - {\bf  {v}} t )$. The stresses at an arbitrary
point of the plane can be obtained by means of the Green function, $G_{ij}
( {\bf  {r}}- {\bf  {r}'} , t - t' )$, with Fourier transform $G_{ij}
( {\bf  {k}} , \omega )$. In the absence of dissipation, the frequency
$\omega$ appears only in combinations of the type  $\omega^2 - c_{L , T}^2
k^2$, where $c_{L , T}$ denote the velocity of longitudinal
and transverse sound waves. Dissipation changes these expressions into
$\omega^2 - c_{L , T}^2 k^2 - i \eta\omega k^2/\rho$.
The Fourier transform of the applied forces can be written as ${\bf  {f}}
( {\bf  {k}},\omega = {\bf  {v} \cdot {k}} )$. Hence, the denominators
in the Green's functions become $( {\bf  {v} \cdot {k}} )^2 - c_{L,T}^2 k^2 
- i \eta k^2{\bf  {v} \cdot {k}}/\rho$.
At low values of $k$, the influence of the viscosity is
negligible. The long distance behavior is well described by the Yoffe solution.
For large $k$ on the
other hand, the viscous term dominates. This term is more anisotropic than the
inertial term, as it contains one power of ${\bf  {v} \cdot {k}}$, instead of
two. Hence, we expect the tendency towards instability to be reduced. 
We can see when this is relevant by putting $k \sim 1/a$, and noting that
large $k$ means
that $k \gg \rho v/\eta$, or equivalently, $\alpha a >>1$. 
For very large $\eta$ the branching threshold should eventually shift. As 
we will see,  we have confirmed this shift using the simulations. 

\section{numerical simulations}

\subsection{Discretization of the Equations of Motion}

In our simulations we work in two dimensions, and discretize the
continuum equations of elasticity by using a spring model
on a triangular two dimensional grid,
following our previous work on quasistatic fracture   
\cite{model1,model2,model3} 
The equation of motion for the displacement, ${\bf u_r}$ of the node at ${\bf 
r}$
combines inertial and viscous terms. In our discrete model, we get the $k^2$ 
dependence
of the attenuation by using the fact that the friction forces can depend
only on the relative velocities of neighboring nodes   \cite{Landau}.
The equations of motion are:
\begin{eqnarray}
m \frac{  \partial^2 {\bf u}_{\bf r}}{\partial t^2} & = & \sum_{\bf r'} K  
{\bf \hat n} \left[{\bf \hat n}\cdot 
( {\bf u}_{\bf r} - {\bf u}_{\bf r'} ) \right] +\nonumber\\ 
&&\sum_{\bf r'} \eta_o 
{\bf \hat n} \left[{\bf \hat n}\cdot 
\left( \frac{\partial {\bf u}_{\bf r}}
{\partial t} - \frac{\partial {\bf u}_{\bf r'}}{\partial t} \right) \right] 
\label{e:motion}
\end{eqnarray}
where the sums in the second term are over the nearest neighbor nodes, 
${\bf r'}$, to node ${\bf r}$ and ${\bf \hat n}$ is the unit
vector from node ${\bf r'}$, to node ${\bf r}$ . 
The fracture process is described by deleting the forces between two
nearest neighbor nodes when the relative strain,
$|{\bf \hat n} \cdot 
\left[{\bf u}_{\bf r} - {\bf u}_{\bf r'}\right]|$, exceeds a threshold, 
$u_{th}$.
This process is irreversible, and the coupling remains zero at all latter
times. The model used here is
deterministic, and the system is always out of equilibrium.

\subsection{The elastic constants of the model}

In order to find the relationship between the parameters of our model
and the experimental constants we need to write down the equation
of motion of our model (Eq. (\ref{e:motion})) in the continuum limit,
\begin{eqnarray}
m \partial_{tt}{\bf u}=Ka^2\left[\frac{3}{8} \nabla^2{\bf u} + \frac{3}{4}
{\bf \nabla} ({\bf \nabla}{\bf u}) \right ]+ \nonumber \\
\eta_0 a^2 \left[\frac{3}{8} \nabla^2\partial_t{\bf u} + 
\frac{3}{4}{\bf \nabla}({\bf \nabla}\partial_t{\bf u}) \right ] \;,
\label{e:discrete}
\end{eqnarray} 
where $a$ is the lattice spacing. Solving the above equations in a finite
difference scheme gives Eq.~(\ref{e:motion}). Then, comparing Eqs.
(\ref{e:continuum}) and (\ref{e:discrete}) one can obtain the relations
we are seeking for. First note that our model gives the following
relations between the continuum parameters,
$\lambda =\mu =3E/8$ and $\psi = \eta$
and a Poisson coefficient of $1/3$. On the other hand,  
\begin{equation}
\frac{3K}{8m}a^2=\frac{3E}{8\rho}
\end{equation}
\noindent where $a$ is the lattice constant of our triangular
spring network. The longitudinal and transverse sound velocities 
are,
\begin{equation}
c_L=\sqrt{3}c_T=\sqrt{\frac{\lambda + 2\mu}{\rho}}=\sqrt{\frac{9K}{8m}}a
\end{equation}  
\noindent From these results the Rayleigh speed can be easily derived
\cite{Landau},
\begin{equation}
c_R=0.9325 c_T
\end{equation}
\noindent Finally, the equations which relate the constants of
our model with the macroscopic parameters of the material are:
\begin{mathletters}
\begin{equation}
m = 3\rho a^2 d/8
\end{equation}  
\begin{equation}
K =8 c_T^2 m/3a^2
\end{equation} 
\begin{equation}
\eta_o = \eta d
\end{equation} 
\end{mathletters}
\noindent where $d$ is the thickness of the sample.
In the next subsection, we discuss
some difficulties in relating $u_{th}$ to macroscopic variables,
particularly due to the discretization scale $a$.

\subsection{Fracture threshold and macroscopic variables}
 
In the following we take $u_{th}=0.1 a$. The process of failure is described
in macroscopic terms by the maximum load above which the material fails.
In our model, the maximum force exerted by the springs, $K u_{th}$, 
should be equal to this load times the lattice spacing, $a$, times the 
thickness of the slab, $d$.  As $K$ is independent of $a$, we find that 
$u_{th} \propto a$.

The description of fracture by discrete element methods always leads to
failure at the smallest scale. In our case, a sample under
load will eventually fail through the snapping of a single row of
springs, and crack widths are always of order $a$\cite{BP98}.
The energy required to create a crack is, because of this effect, 
dependent on the discretization, and goes to zero as the lattice spacing
is decreased. In the present case, the energy needed to create
a crack of (macroscopic) length $l$ goes as $\frac{l}{a} K
u_{th}^2 \propto a$. Thus, we cannot fit the tensile strength
and the fracture energy at the same time in a discretization
independent way. This drawback can be inferred from the
existence of a characteristic length which combines these quantities,
as analyzed in the introduction. 

Macroscopic crack energies can be obtained from models which incorporate
non local effects\cite{PEG94}.  
However, detailed microscopic simulations\cite{Abraham} show cracks
of atomic width, with little or no damage outside a zone of
microscopic dimensions. They are also difficult to reconcile with
the existence of crack energies with macroscopic values. The origin of
macroscopic failure zones in quasibrittle materials such as PMMA
is not well understood.

\subsection{The dissipation term}

Qualitatively, each node represents a region in the material comparable
to the scales relevant to the experimental situation. In our case, we have
$l_{ch} \approx 60$ microns. We use  a phenomenological damping term,
$\eta_o \sim 1$  in units where $K=m=1$, which implies that
sound waves at this length scale are in the overdamped regime.
 As mentioned earlier, the sound attenuation
in glassy polymers has a complicated dependency on frequency.
Our choice of $\eta$ overestimates the experimental value in the
GHz range\cite{Maris}, although probably underestimates it at
lower frequencies (note that our model assumes that $\alpha \Lambda \sim
\Lambda^{-1}$ at all wavelengths).  
The scheme used here is intermediate between a full scale atomic
simulation  \cite{Abraham}, and more phenomenological models  \cite{Marder},
where dissipation takes place within the units used in the discretization.

\subsection{Drawbacks of the model}
As mentioned earlier, the main difficulty with the model is the fact
that the crack energy does not scale with the level of discretization.
Only for a given lattice constant, $a$, the crack energy and the maximum
load can be consistent with macroscopically determined values.
On the other hand, instability criteria based on energy considerations,
such as the Griffith criterion and extensions thereof, depend only on energy
differences. Hence, we do not think that the problem discussed here
is a serious obstacle to the analysis of the instabilities of moving
cracks. As there are substantial differences between discrete
and continuum models\cite{KL98}, it would be interesting to
analyze further the relevance of the intrinsic length scale determined by
the macroscopic parameters which describe fracture. 

\subsection{Numerical Procedures}

Simulations have been performed in rectangular strips with the y orientation
along one of the axis of the triangular lattice. The lattices shown in the
figures are 50 nodes wide and 275 nodes long. 
The boundary condition is fixed displacement of the edges so that
the initial strain is below $u_{th}$.
Then, bonds near the lower horizontal edge
are broken at a fixed rate, so that the velocity of the crack is well
below $c_T$. To integrate the equations of motion we use Heun's method with a
time step small enough for appreciate small variations in the velocity of the
crack. Once the crack is sufficiently long, strains near its tip
begin to exceed $u_{th}$, and the crack continues growing by itself.
Very short cracks do not grow on their own, because the strains
at the tip do not exceed $u_{th}$. The minimum size for self-sustained
growth decreases with increasing dissipation.

\section{Results}

\subsection{Stable and unstable cracks}
\begin{figure} 
\epsfxsize=\hsize 
\centerline{\epsfbox{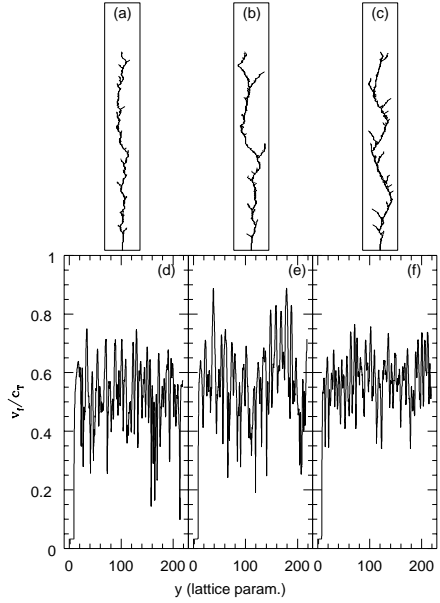}}
\caption{Behavior of inertial cracks when the external strain, $u_{appl}$, is
varied. (a) $u_{appl}$=0.024. (b) $u_{appl}$=0.026. (c) $u_{appl}$=0.028. 
And also their
velocity (in units of $c_T$) as function of position of the advancing crack
tip. (d) $u_{appl}$=0.024. (e) $u_{appl}$=0.026. (f) $u_{appl}$=0.028. 
The threshold for breaking is $u_{th}=0.1$.}
\label{f:inertial}
\end{figure}

In the absence of damping, straight cracks become unstable on short time
scales. Typical results for cracks growing in a narrow slab under an applied
strain at the edges are shown in Fig.~\ref{f:inertial} (a)-(c). The crack tips
accelerate, exactly as predicted in Eq. (\ref{eqacc}) until they approach
$c_Y$, and then they branch. The velocity of the upper-most part of the crack
pattern is depicted in Fig.~\ref{f:inertial} (d)-(f). We note that the crack
velocity strongly oscillates as a function of the crack length. 

\begin{figure} 
\vspace{0.5cm}
\epsfxsize=\hsize 
\centerline{\epsfbox{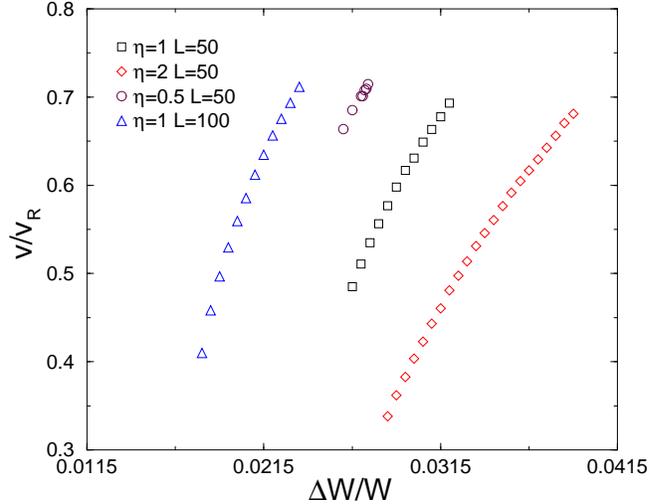}} \vspace{-5cm}
\caption{Terminal velocity for a stable crack in units of the Rayleigh speed
as function of the applied strain for different values of the viscosity and
width of the system.}
\label{f:velos}
\end{figure}

We find that straight cracks can be stable (cf Fig.~\ref{f:ruido}a, below), 
and move at constant velocity, 
in the presence of dissipation. As the driving force is increased, we observe
a branching instability. 
This behavior is what is predicted by Eq. (\ref{eqterm}). 
If the terminal velocity
is below $c_Y$ (which we assume to be independent of $\eta$, see below) the 
crack will be slowed down and prevented from branching.
This behavior is shown in Fig.~\ref{f:velos} where the terminal stable velocity
in units of $c_R$ is plotted against the displacement at the borders of the
system. The evolution to increasing velocity proceeds as the external
displacement is increased. The curves end at the branching instability. From
them, we can deduce that this threshold is independent of the parameters of the
simulation (size and viscosity) and $\approx 0.7c_R$, in agreement with
Yoffe~\cite{Yoffe} calculations. 

The reason for the insensitivity of the branching threshold
to $\eta$ was given above. We find in our simulations
that to shift the threshold significantly
below 0.7$c_{R}$, $\eta$ must be greater than 7 for a 50 $\times$ 300 mesh,
which is, we think, an unphysically large value. 
Nevertheless, the branching threshold reported in 
\cite{Ciliberto2} $\approx 0.45c_R$,
is around $35\%$ smaller than in the numerical 
simulations.

The fact that we do not see an abrupt change in the velocity before the
branching threshold is related with the method we use to perform the
simulations. Boudet and Ciliberto~\cite{Boudet1} demonstrated experimentally
that this jump was not present when sound was added into the system. This is
what we do in the simulations: the slow cutting of bonds that initiate the
fast failure is a source of sound into the system.


\begin{figure} 
\mbox{\psfig{file=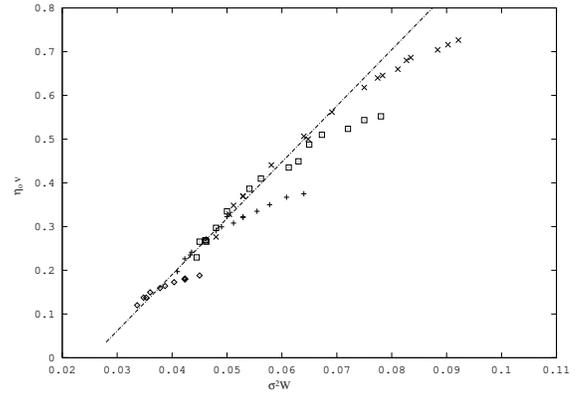,height=2.2in,angle=270}}
\caption{Data collapse of $\eta_o v$ plotted against $\epsilon^2 W$. 
$\diamond, \eta_o = 0.5$; +, $\eta_o=1.0$; $\Box , \eta_o=1.5$; {\sf X}, 
$\eta_o=2.0$, for $12 < W < 50$ and $0.029 < \epsilon < 0.058$.The straight
line is fitted to all the points for which $v<0.3$.}
\label{f:escalado}  
\end{figure}

We can directly verify the validity of Eq. (\ref{eqterm}) by considering
a number of different sets of the parameters $\epsilon, \eta, W$, such as
those shown in Fig.~\ref{f:velos} and viewing 
our data in the form of a data collapse. This is done in Fig.~\ref{f:escalado}
where we show that $\eta v$ is very accurately a linear function of
$\epsilon^2 W$ for low speeds. For larger speeds, of the order of $c_Y$ there
are deviations from Eq.~\ref{eqterm}, as we expect.

\subsection{Thermal noise and the branching threshold}
Numerically is possible to obtain lower critical branching velocities by adding
a random noise in Eq.~\ref{e:discrete}, as will be discussed in detail
in a forthcoming publication \cite{noise}. This random noise 
is in the form that
it has zero mean at its correlation is
\begin{equation}
<{\bf \gamma}_i\cdot {\bf \gamma}_j>=\xi T,
\end{equation}
when $i=j$, and zero otherwise. $T$ is the temperature, and $\xi$ is a 
parameter that controls numerically the amplitude of the
noise. In Ref.~ \cite{noise} $\xi$ is related to $\eta$ using the
fluctuation-dissipation theorem. Here we take a more phenomenological
approach in order to simply illustrate the effect.

\begin{figure} 
\epsfxsize=1.4\hsize 
\centerline{\epsfbox{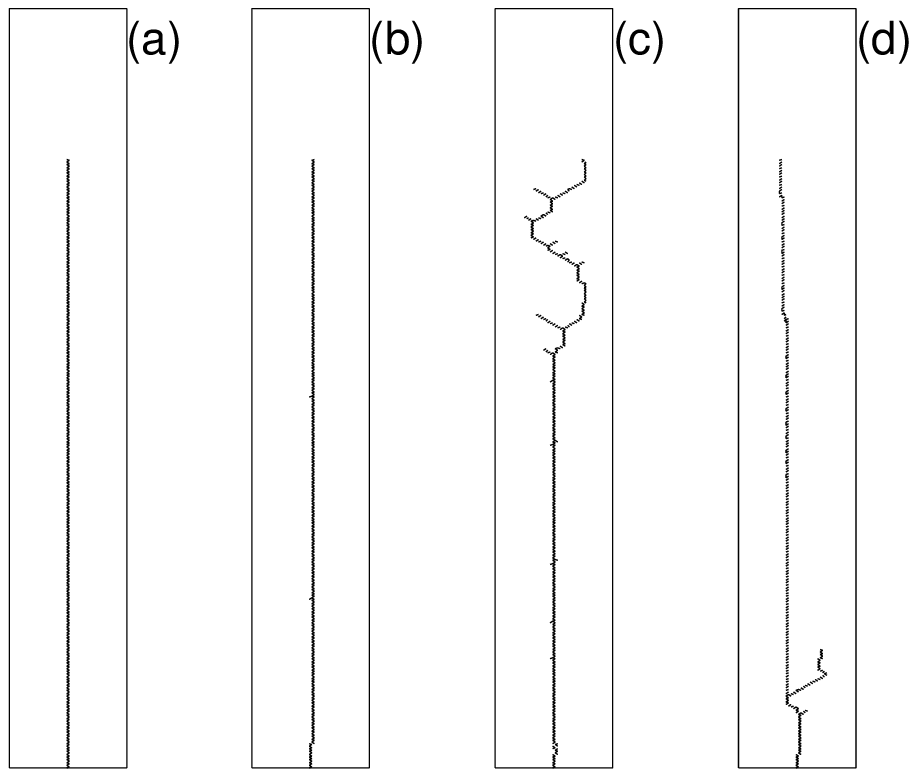}} \vspace{-7cm}
\epsfxsize=\hsize 
\centerline{\hspace{1.5cm}\epsfbox{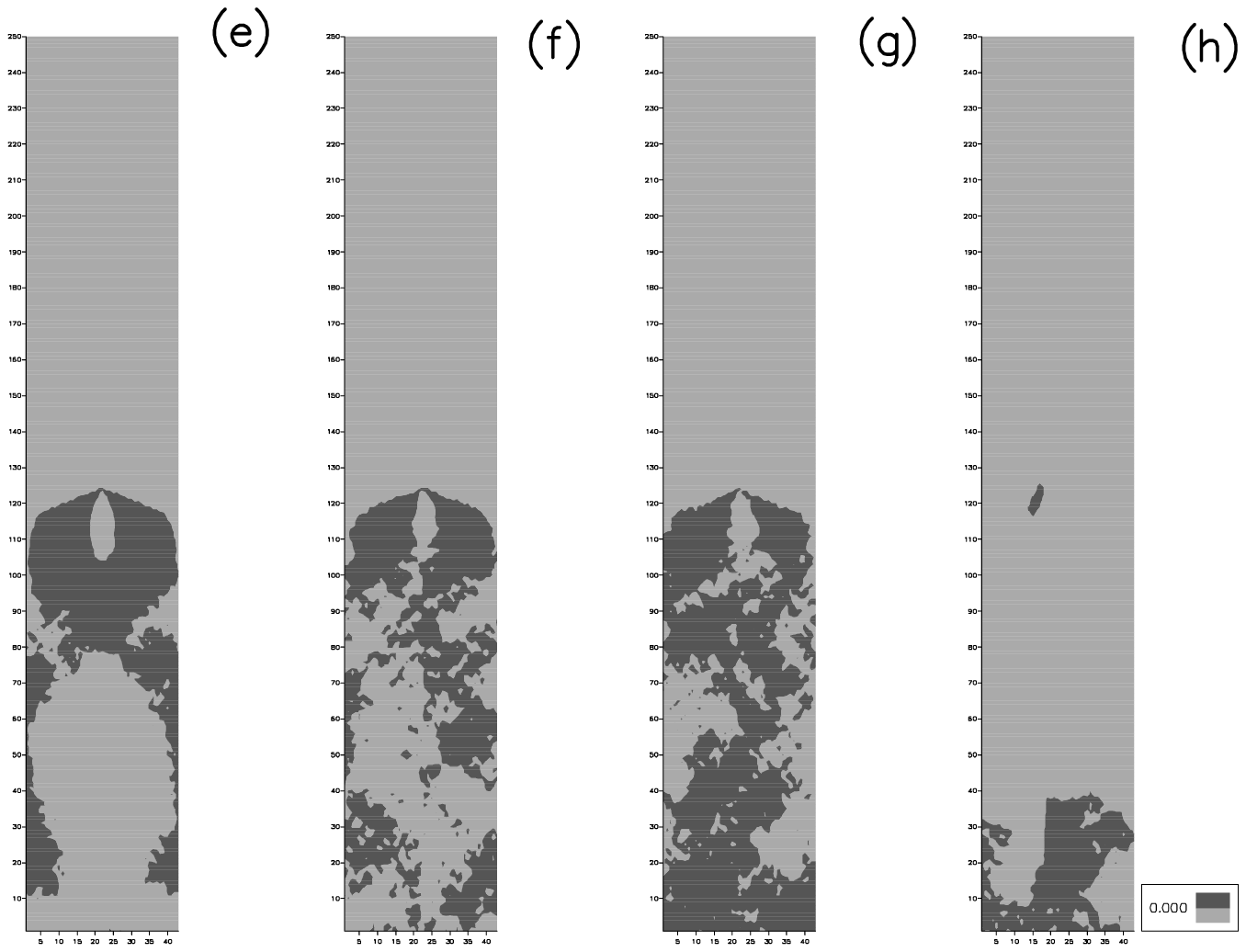}} \vspace{-6cm}
\caption{Shapes of the cracks when noise is added (a) $\xi=10^{-9}$, (b)
$\xi=4\times 10^{-8}$, (c) $\xi=10^{-7}$, and (d) $\xi=2\times 10^{-7}$. The
lower figures (e-h) show the sign of the Coterell and Rice parameter 
($\sigma_{yy}-\sigma_{xx}$): dark grey positive and light grey negative. 
Stresses are taken when the crack has advanced half the length of
the system. The noise is the one of the corresponding above figure (a-d).}
\label{f:ruido}
\end{figure}

Let us take one of the stable cracks whose
velocity is plotted in Fig~\ref{f:velos}. The one with 
$\Delta W/W$=0.03, 
$\eta$=1, and $L$=50 has a velocity of $\approx 0.65c_R$. Fig.~\ref{f:ruido}
show the shape of the cracks for different values of $\xi$. The temperature
is set to a fixed arbitrary value of 100. When the noise is very low 
($10^{-9}$) nothing happens to the crack. When it becomes higher, always at
the same terminal velocity, small side branches like in the experiments can be
obtained. The spacing between branches decreases as the noise intensity is
increased, and, in some cases (see Fig.~\ref{f:ruido}(c)), the crack can
disestabilize at later stages in the growth process, until finally the crack
becomes unstable (Fig.~\ref{f:ruido}(d)).

Since in our simulations we have all the information of the displacements at
all points of the network, we can compute the stress tensor at any time of
the evolution of the crack. In particular it is interesting to see the values
of $\sigma_{yy}-\sigma_{xx}$ around the crack tip. According to Coterell and
Rice~\cite{CR}, this quantity describes the stability of the cracks. It should
be less than zero at the crack tip for an stable crack and greater than zero
when it becomes unstable. This parameter is shown in the lower part of
Fig.~\ref{f:ruido} when the corresponding cracks in the upper part of the
figure have advanced half the length of the system. The three cracks with
lower noise intensity are, according to this criterion, still stable, but the
one with higher noise has a tendency to continue deviating to the left.

This method for representing the stress tensor with a wide variety of
parameters in an elastic medium can provide a valuable tool for inspecting
the conditions for the stability, and the analytical approximations that can
be made~\cite{futuro}. 

\section{Heating and energy dissipation}

In the previous section, examples of crack propagation in the
presence of thermal noise due to an external environment,
 have been discussed. However, the local
temperature around the crack tip must also take into account the
energy dissipated by the viscosity term that we have in the
equations.
Near the crack tip typical deviations of the nodes from equilibrium are of
order $a$. Typical velocities are of order $K u_{th}/\eta_o$. The energy
dissipated per node and per unit time is $\sim K^2 u_{th}^2/\eta_o$.
In terms of macroscopic quantities, the energy generated per unit time and
unit volume is $\sim \rho^2 c^4 a^2 \sigma^2_c/(\mu E^2)$, where $c$ is
some average of the longitudinal and transverse sound velocities, $E$ is the
Young's modulus and $\sigma_c$ is the macroscopic elastic limit. This
dissipation generates thermal gradients. They will be determined by the
condition:
\begin{equation}
\frac{\partial T}{\partial t} = \kappa \nabla^2 T
+ \frac{1}{\rho c_e} \frac{\partial {\cal E}}{\partial t} = 0
\end{equation}
where $\kappa$ is the thermal diffusion coefficient, 
${\cal E}$ is the energy being dissipated
and $c_e$ is the specific
heat. Assuming that most of the dissipation takes place at distances from the
crack tip comparable to its radius, we find that the temperature increase at
the tip can be written as:
$\Delta T_{tip} \sim c^4 a^4 \sigma_c^2/(\kappa \mu c_e E^2)$
This expression is highly sensitive to the choice of $a$, the tip radius.
Hence, it is difficult to make accurate estimates of the expected heating. 
Experimentally, significant increases in temperature near the crack tip have
been reported  \cite{heating}. Energy dissipation has also been observed 
in  \cite{Fineberg1}, where it is argued that most of the energy is spent in
increasing the crack surface. However, even for slow, straight cracks, a
significant rise in energy dissipation as function of velocity is reported.
In our simulations the elastic energy lost when one spring is cut goes
into surface energy, whereas the viscous dissipation goes into heat.
Heating of the crack tip increases
thermal noise there. This could be quite significant since, near, but below
the branching speed the stress distribution becomes nearly isotropic, so that
relatively small thermal effects could lead to branching. The considerations
in this section will be worked out in detail in Ref.~\cite{noise}.

\section{Conclusions}

The analysis reported here indicates that viscous effects change
significantly the propagation, and instabilities, of cracks in
brittle materials.
The general features that we have found should be reproduced, for
example, in PMMA, even though the viscosity is a more
complicated function of frequency than the one considered here.
Some of the characteristics of the experimental results
\cite{Swinney,Ciliberto1,Fineberg1,Ciliberto2} are already 
qualitatively described by the present approach.
In particular, our approach would explain why experiments in glass are harder
to perform than in PMMA: its associated viscosities are lower than in PMMA
and thus it is closer to the instability.
On the other hand, the branching threshold seems to be lower in
the experiments than in our numerical simulations for the chosen parameters.
This fact has also been discussed and shown to be due to other effects not
contained within the model, but that can be incorporated as external noise.
Of course, the richness and complexity of fracture in these
materials will require further investigations. We hope that
the approach herewith proposed will help to improve our
understanding of these interesting phenomena.

\section{Acknowledgments.}

We acknowledge many fruitful discussions with
J. Planas, J. Colmenero, F. Abrahams, 
P. Espa\~nol and M. A. Rubio. We are also grateful to S. Ciliberto and J.-F.
Boudet for a most illuminating description of the experimental situation.
FG is supported by CICyT grant no. PB96-0875, EL by
CICyT grant no. PB96-0085, and LMS and SVG by NSF grant DMR 94-20335. LMS
also acknowledges help from the Iberdrola Foundation.

\section{Appendix}

The energy dissipated by sound waves goes into other excitations
of the system. Taking the sound
velocity of PMMA 
to be $v_s \approx 10^5$ cm/s, a mode of wavelength of one micron
has a frequency of $10^9$Hz. In energy units, it corresponds to $\approx
7 \times 10^{-7}$eV.  At sufficiently low temperatures,
the main source of inelastic scattering at these
frequencies, as seen by neutron scattering, is  quantum tunneling
between equivalent configurations of the CH$_3$ groups attached to the
polymer\cite{Cetal98}. The possibility that these excitations play a role
in sound attenuation was suggested in\cite{WE78}.

An acoustical phonon modulates the distance between polymers. At the low
frequencies involved, the backbone of the polymer cannot be excited, and can
be considered rigid. The motion of a nearby polymer changes the potential
acting on a given CH$_3$ group, breaking the initial threefold symmetry.
The asymmetry in the potential induces transitions between the
quantum levels of the CH$_3$ group, and leads to dissipation.

The interaction between the neutral CH$_3$ unit and other parts of the
polymer arise from mutual induced polarization. Assuming that neither
part has a finite electric dipole, the interaction energy is
given by the van der Waals expression:
\begin{equation}
E = \frac{e^4}{\epsilon^2 r^6} \sum_{m,n} \frac{| \vec{d}^a_m \vec{d}^b_n - 3
( \vec{d}^a_m \vec{r} ) ( \vec{d}^b_n \vec{r} ) / r^2 |^2}{
\Delta_m + \Delta_n}
\label{vdw}
\end{equation}
where $r$ is the distance, $\epsilon$ is the dielectric constant
due to the rest of the material, $\vec{d}^a_m = \langle 0 | \vec{r} | m \rangle$
represents a matrix element of $\vec{r}$ between states of unit $a$
( and a corresponding expression for $\vec{d}^b_n$), and $\Delta_m$
is the energy difference between the ground state of unit $a$ and a given
excited state.

The order of magnitude of $E$ in (\ref{vdw}) is
$E \sim \frac{e^4 d^4}{\epsilon^2 
r^6 \Delta}$, where $d$ goes as the dimension of 
the unit, $\Delta$ is a typical electronic excitation energy, and $r$ 
is the separation.

The splitting between configurations of the CH$_3$ unit goes as the difference
in the interaction energy (\ref{vdw}) at nearby H sites.
Hence, it goes as $\frac{\partial E}{\partial r} d_{H-H}$.

A phonon which induces displacements $\vec{u}_k$ in a given molecule
leads to a change in the intermolecule distance $r$ of order $(
\vec{k} \vec{r} ) \vec{u}_k$. Using the golden rule, the energy per unit
time absorbed by a given CH$_3$ group goes as:
\begin{equation}
\frac{\partial E}{\partial t} = \frac{1}{\hbar} \left( \frac{e^4 d^4}
{\epsilon^2 r^6 \Delta} \right)^2 \frac{d_{H-H}^2}{r^2} k^2 u_k^2 
\omega_{ph} \rho_{tunn} ( \omega_{ph} )
\label{diss}
\end{equation}
where $\rho_{tunn} ( \omega )$ is the density of tunneling centers of
energy splitting $\omega$. 
The energy dissipated per unit volume
can be estimated from (\ref{diss}) by multiplying that expression by
the number of CH$_3$ groups per molecule, $N$, and dividing by the volume
of the molecule, $\Omega$. 

On the other hand, the energy dissipated per unit volume, can be
written as\cite{Landau}:
\begin{equation}
\frac{\partial E}{\partial t} \propto \eta \omega_{ph}^2 k^2 u_k^2
\label{macro}
\end{equation}
where $\eta$ stands for an average of the macroscopic viscosity.
In terms of $\eta$, the sound attenuation goes as
$\frac{\eta \omega_{ph}^2}{\rho c^3}$, where $\rho$ is the density,
and $c$ is the sound velocity. Making use of the microscopic parameters,
this gives:
\begin{equation}
\alpha \sim k \frac{N}{\hbar \rho \Omega c^2} \frac{d_{H-H}^2}{r^2}
\left( \frac{e^4 d^4}{\epsilon^2 r^6 \Delta} \right)^2 \rho_{tunn} ( \omega_{ph} 
)
\label{attenuation}
\end{equation}
This expression is very sensitive to the value of $r$, the distance between
nearby chains. A rough estimate can be made by assuming
$\frac{\rho \Omega c^2}{N} \sim \Delta \sim 1$eV, $d_{H-H} \sim d \sim
1$\AA, $r \sim 5$\AA , $\rho ( \omega ) \sim ( 1 \mu {\rm eV} )^{-1}$ 
and $\epsilon = 1$. Using these parameters, we obtain
$\alpha \Lambda \sim 1 - 10$, where $\Lambda$ is the wavelength of
the phonon. The dependence on frequency of $\alpha \Lambda$ is that
of $\rho_{tunn} ( \omega )$, which, in the relevant range of frequencies,
$\sim 1 \mu$eV, is roughly constant\cite{Cetal98}. This gives
$\alpha \propto \omega$, in line with 
\cite{Maris}. Below lower frequencies, $\sim 1$kHz,
$\rho_{tunn} ( \omega ) \propto \omega$, leading to $\alpha \Lambda \propto
\Lambda^{-1}$, and $\alpha \propto \omega^2$, similar to the behavior 
reported in\cite{Mark}. 

Quantum tunneling is suppressed by thermal fluctuations, which break
the degeneracy of the three potential minima seen by the CH$_3$ groups.
We assume that these fluctuations arise from changes in the van der Waals
interactions with neighboring polymers. The presence of a lattice vibration,
of momentum $k$ and amplitude $u_k$ induces a splitting between 
equivalent minima of:
\begin{equation}
\Delta E \sim \frac{e^4 d^4}{\epsilon^2 r^6 \Delta} \frac{d_{H-H}}{r}
k u_k
\label{splitting}
\end{equation}
The derivation of this expression follows the analysis presented earlier.
At finite temperatures, $u_k$ shows random fluctuations.
The mean square deviation of $\Delta E$ is given by:
\begin{equation}
\langle ( \Delta E )^2 \rangle \sim \left( \frac{e^4 d^4}{\epsilon^2 r^6 \Delta}
\right)^2 \frac{d_{H-H}^2}{r^2} \langle k^2 u_k^2 \rangle
\label{fluctuations}
\end{equation}
and:
\begin{equation}
\langle k^2 u_k^2 \rangle \sim \Omega \int^{\hbar c k \ll k_B T}
d^3 k \frac{k^2 k_B T}{\rho \Omega \hbar^2 c^2 k^2} \sim
\frac{k_B T}{\Omega \rho c^2} \left( \frac{T}{\omega_D} \right)^3
\label{thermal}
\end{equation}
where $\omega_D$ is the Debye temperature. In terms of the attenuation rate
calculated above, we can write:
\begin{equation}
\langle ( \Delta E )^2 \rangle \sim \frac{\alpha \Lambda}{N}
 \frac{k_B T}{\rho_{tunn}
( E )} \left( \frac{T}{\omega_D} \right)^3
\label{fluctuations2}
\end{equation}
Using the same parameters as above, and setting $\omega_D = 300$K, we
find that the fluctuations $\Delta E$ are of order $1 \mu$eV when $T 
\sim 10$K.

At higher temperatures, the dynamics of CH$_3$ groups lose
coherence. They still influence sound attenuation, because they 
mediate interactions between phonons.

Note, finally, that the mechanism discussed here does not lead to 
plastic effects. The material remains brittle at the scales
of interest, as it seems to be the case in PMMA. 
Other mechanisms, such as the irreversible relaxation of
structural defects, may lead to viscoplastic effects\cite{FL98}.


\end{document}